\documentclass[journal]{IEEEtran}
\usepackage{amsmath}
\usepackage{graphicx}
\usepackage{enumerate}

\usepackage{url} 

\newcommand{\nn}{\nonumber \\}

\newcommand{\col}{\textrm{col}}
\newcommand{\rvec}{\textrm{vec}}

\newcommand{\var}{\textrm{var}}

\newcommand{\cA}{\mathcal{A}}
\newcommand{\cB}{\mathcal{B}}

\begin{document}

\title{Traffic Rate Network Tomography with Higher-Order Cumulants}
\author{
 Hanoch Lev-Ari, \IEEEmembership{Fellow,~IEEE},
  Yariv Ephraim  \IEEEmembership{Fellow,~IEEE},
 Brian L. Mark, \IEEEmembership{Senior Member,~IEEE}

\thanks{This work was supported in part by the U.S.\ National
  Science Foundation under Grant CCF-1717033.}
  
\thanks{H.\ Lev-Ari is with the Dept.\ of Electrical and Computer Engineering, Northeastern University, Boston, MA 02115 (e-mail: levari@ece.neu.edu).}  
  
  \thanks{Y.\ Ephraim and B.L.\ Mark are with the 
    Dept.\ of Electrical and Computer
    Engineering, George Mason University, Fairfax, VA 22030 
    (e-mail: yephraim@gmu.edu, bmark@gmu.edu).}
 
}

\maketitle

\begin{abstract}
Network tomography aims at estimating source-destination traffic rates from link traffic 
measurements. This inverse problem was formulated by Vardi in 1996 for Poisson traffic 
over networks operating under deterministic as well as random routing regimes. In this 
paper we expand Vardi's second-order moment matching rate estimation approach to higher-order 
cumulant matching with the goal of increasing the column rank of the mapping and consequently improving 
the rate estimation accuracy. We develop a systematic set of linear cumulant 
matching equations and express them compactly in terms of the Khatri-Rao product. 
Both least squares estimation and iterative minimum I-divergence estimation are considered. 
We develop an upper bound on the mean squared error (MSE) in least squares rate estimation 
from empirical cumulants. We demonstrate for the NSFnet that 
supplementing Vardi's approach with third-order empirical 
cumulant reduces its averaged normalized MSE relative to the theoretical 
minimum of the second-order moment matching approach by   
about $12\%-18\%$. This minimum MSE  
is obtained when Vardi's second-order moment matching approach is based on the theoretical 
rather than the empirical moments. 
\end{abstract}

\begin{IEEEkeywords}
Network traffic, network tomography,  inverse problem, higher-order cumulants.
\end{IEEEkeywords}

\section{Introduction} \label{introduction}

Network tomography was formulated by Y. Vardi in the seminal paper \cite{Vardi:1996}.
The goal is to estimate the rates of traffic flows over source-destination pairs   
from traffic flows over links of the networks. Traffic flow rates are  
measured, for example, by the number of packets per second. We use $X$ to denote a vector 
of $L$ source-destination traffic flows, and $Y$ to denote a vector 
of $M$ link traffic flows. Normally, $M \ll L$. All source-destination traffic 
flows are assumed independent Poisson random variables. Thus, $X$ comprises a vector  
of $L$ independent Poisson random variables. Link traffic flow measurements are assumed passive and 
require no probes. For networks operating under a deterministic routing regime, traffic 
flows from each source node is routed to the destination node over a fixed path. Traffic flows over 
each link may originate from multiple source-destination traffic flows. Thus, we 
define a binary variable $a_{ij}$ such that $a_{ij}=1$ when traffic over source-destination 
$j$ passes through link $i$ and $a_{ij}=0$ otherwise. 
 We also define an $M \times L$ binary 
routing matrix $A=\{a_{ij}, i=1,\ldots,M; j=1,\ldots,L\}$ which is assumed known 
throughout the paper. It follows that $Y=AX$. The expected value of the $j$th 
Poisson source-destination traffic flow 
constitutes its traffic flow rate. Thus, the  
vector of source-destination traffic flow rates is the expected 
value of $X$ 
which we denote by $\lambda = E\{X\}$. 

The network tomography problem is that of estimating the rate vector $\lambda$ from 
given realizations of $Y$. This is a rather challenging inverse problem since 
$Y=AX$ is an underdetermined set of equations, and the estimate of $\lambda$ must be 
non-negative. Vardi invoked the terminology ``LININPOS'' for (LINear INverse POSitive 
problem). Maximum likelihood estimation of $\lambda$ is not feasible since the 
components of $Y$ are dependent Poisson random variables with no explicitly known 
distribution. The expectation-maximization (EM) algorithm is also not useful for this 
problem as it requires calculation of the conditional mean $E\{X \mid Y\}$ which 
in turns requires knowledge of the joint distribution of $X$ and $Y$.  Vanderbei and 
Iannone \cite{Vanderbei:1994} developed an EM algorithm in which $E\{X \mid Y\}$ is simulated.
A common approach is to rely on the explicit form of $E\{X \mid Y\}$ for jointly 
Gaussian $X$ and $Y$, see, e.g.,  \cite{Vardi:1996}, \cite{Cao:2000}, \cite{Liang:2003}.  
 
Vardi resorted to moment matching in which $\lambda$ is estimated from a linear matrix equation 
relating the first two empirical moments of $Y$ to the corresponding first two theoretical 
moments of $AX$. The approach is applicable to networks operating under deterministic as well 
as random routing regimes. In the 
latter case, there are multiple alternative paths for each source-destination pair which can be 
selected according to some probability law. 
Vardi invoked an iterative procedure for estimating $\lambda$ which was previously  
developed for image deblurring \cite{Richardson:1972}. This useful procedure will be 
discussed in Section \ref{iDiv}. 
Vardi's work ignited extensive research in the areas of network inference and medical tomography.

In this paper we explore the benefits of supplementing Vardi's second-order moment matching approach 
with higher-order empirical cumulants. Here too the cumulant matching equation is linear in $\lambda$, and the 
approach is applicable to networks operating under deterministic as well as random routing regimes. 
Higher-order cumulants introduce new useful information on the unavailable distribution of $Y$ which 
can be leveraged in estimating the source-destination flow rates. 
By using a sufficient number of empirical cumulants, the linear mapping involved 
in the cumulant matching approach can achieve full column rank. In an ideal scenario where 
the cumulant matching equations are consistent, and the cumulants of the link traffic flows 
are accurately known, the rate vector can be recovered without error. As we demonstrate in Section 
\ref{num}, this ideal situation is approachable when a sufficiently 
large number of realizations of $Y$ are available.

A vast literature exists on network tomography. 
Several forms of network tomography have been studied in the literature depending on the
type of measurement data and the network performance parameters (e.g., rate, delay) of interest:
(i) source-destination path-level traffic rate estimation based on link-level traffic 
measurements~\cite{Vardi:1996,Tebaldi:1998,Cao:2000,Liang:2003,Mardani:2016, Singhal:2007, Hazelton:2015}; 
(ii) link-level parameter estimation based on end-to-end path-level traffic
measurements~\cite{Vanderbei:1994}, \cite{Caceres:1999,LoPresti:2002,Tsang:2003,Ma:2013,Etemadi:2017};  
(iii) network topology discovery from traffic 
measurements~\cite{Jin:2006,Eriksson:2010,Ni:2010,Anandkumar:2011}.  Approaches
to network tomography can also be characterized as 
active (e.g., \cite{Caceres:1999,LoPresti:2002,Kumar:2006,Rabbat:2006}), whereby explicit
control traffic is injected into the network to extract
measurements, or passive (e.g., \cite{Vardi:1996,Cao:2000,Yao:2012,Etemadi:2017}), 
whereby the observation data is obtained from existing network traffic. 
A somewhat outdated survey circa 2003 can be found in~\cite{Castro:2004}.  
A more recent survey~\cite{Qin:2014} discusses network tomography in conjunction with network coding.

Closest to Vardi's work is the contemporary work of Vanderbei and Iannone \cite{Vanderbei:1994} 
which relies on a Poisson model for incoming traffic. The goal is to estimate  
the rate of traffic on each link connecting input and output nodes from traffic counts at the 
input and output nodes. 
Vanderbei and Iannone did not resort 
to moment matching but rather developed a simulation based EM algorithm for maximum likelihood estimation of the 
rates. A thorough Bayesian approach to Vardi's problem was developed by Tebaldi and West \cite{Tebaldi:1998}  
using Markov Chain Monte Carlo simulation. See also \cite{Hazelton:2015}.  
Another closely related work to Vardi's problem appeared in \cite{Cao:2000}, \cite{Liang:2003}, \cite{Castro:2004} 
where maximum likelihood estimation of the 
source-destination rates from link data was implemented under a Gaussian rather than a Poisson traffic model. 
 In \cite{Mardani:2016}, source-destination rates were 
estimated by utilizing spatiotemporal correlation of nominal traffic, and the fact that traffic anomalies
are sparse. In \cite{Singhal:2007}, conditions for identifiability of higher order cumulants in estimation 
of source-destination traffic from link measurements were established. In \cite{Ma:2013} an algorithm was 
developed for choosing a set of linearly 
independent source-destination measurement paths from which additive link metrics are individually estimated.  
 
The plan for this paper is as follows. In Section \ref{iDiv} we present the 
minimum I-divergence iterative procedure which plays an important role in 
this paper. In Section \ref{main} we address rate estimation 
in networks with deterministic routing. 
We develop a new systematic set of cumulant matching equations for estimating the rate vector 
from up to the fourth-order empirical cumulant. We also develop an upper bound on the mean squared error (MSE) in least squares 
estimation of the rate vector from empirical cumulants. We conclude this section with a  
discussion on some implementation issues of the proposed cumulant matching approach. Complexity
of the approach is discussed in Section \ref{complex}. In Section 
\ref{rand} we discuss rate estimation in networks operating under a random routing regime. 
Numerical results for the NSFnet \cite{Bonani:2016} are presented in 
Section \ref{num}.  Concluding remarks are given in Section \ref{comm}. 
Technical details of the derivations are deferred to the Appendix. 

\section{Minimum I-Divergence Iteration} \label{iDiv}

Under the Poisson model for $X$, the moment (and cumulant) matching  
approach results in a set of linear equations in $\lambda$. 
Generally speaking, the set of equations 
has the form of $\hat{\eta}(Y)=B \lambda$ 
where $\hat{\eta}(Y)$ is a vector of $M_b$ empirical  moments (first and second-order in \cite{Vardi:1996})  of 
$Y$, $B=\{b_{ij}, i=1,\ldots,M_b; j=1,\ldots,L\}$ is a constant zero-one matrix that depends on $A$ 
but not on $\lambda$, and $B \lambda$ represents the corresponding theoretical moments.  

To estimate $\lambda$ that satisfies $\hat{\eta}(Y)=B \lambda$, 
Vardi proposed to use an iterative estimation approach which originated in the study of another inverse 
problem concerning image deblurring \cite{Richardson:1972}, \cite{Snyder:1992}. 
Let $\lambda_j^{\textrm{old}}$ denote a current estimate of the $j$th component of 
$\lambda$, and let $\lambda_j^{\textrm{new}}$ denote  the new estimate of that component 
at the conclusion of the iteration. Let $(B \lambda^{\textrm{old}})_i$ denote the 
$i$th component of $B \lambda^{\textrm{old}}$. Similarly, let $\hat{\eta}_i(Y)$ 
denote the $i$th component of $\hat{\eta}(Y)$. The iteration is given by 
\begin{align}\label{e77}
\lambda_j^{\textrm{new}} &=  \lambda_j^{\textrm{old}} 
\sum_{i=1}^{M_b} \bar{b}_{ij}
\frac{\hat{\eta}_i(Y)}{(B \lambda^{\textrm{old}})_i} \textrm{ where }  
\bar{b}_{ij} :=\frac{b_{ij}}{\sum_{t=1}^{M_b} b_{tj}},  
\end{align}
for $j  =  1,\ldots,L$.
This iteration is particularly suitable for solving positive inverse problems. 
It reaches a fixed point when the moment matching equation $\hat{\eta}(Y)=B \lambda^{\textrm{old}}$
is satisfied. 
The iteration was studied by Snyder, Schulz and O'Sullivan \cite{Snyder:1992} in a 
similarly formulated application of image deblurring. It was shown to monotonically decrease 
Csisz\'{a}r's I-divergence \cite{Csiszar:1991} between the original image convolved with the kernel, 
and the observed blurred image. The procedure 
turned out to be an EM iteration in the positron emission tomography problem, which  
follows a similar formulation as network tomography, but with the crucially facilitating 
difference that the Poisson components of $Y$ are now \textit{independent} random variables
\cite{Shepp:1982}, \cite{Vardi:1996:1}. 
This iteration will play a central role in our cumulant matching approach.

\section{Cumulant Matching in Deterministic Routing} \label{main}

In this section we present our cumulant matching approach and its theoretical performance bound
for networks operating under a deterministic routing regime. 
The goal is to estimate the source-destination rate vector $\lambda$  
from $N$ realizations of the link traffic flow $Y$ where 
$Y =A X$ and the routing matrix $A$ is known. Conditions for identifiability of $\lambda$ were given in \cite{Vardi:1996}. 
Specifically, the parameter $\lambda$ is identifiable if all columns $\{a_1, \ldots,a_L\}$ 
of $A$ are distinct 
and none is null.

\subsection{Cumulant Matching}  
 
Let $\bar{X} :=X-E\{X\}$ where $X$ is any real random vector with finite mean and possibly 
dependent components. 
The vectorized $k$th order central moment of $X$ is given by  
\begin{align}\label{b15}
\mu_k(X) = E\{\underset{k \textrm{ times }}{\underbrace{\bar{X} \otimes \bar{X} \otimes \cdots \otimes \bar{X}}}\} 
\end{align}
where $k$ is any positive integer and $\otimes$ denotes the 
Kronecker product. 
When the length of $X$ is $L$, the 
length of the vector $\mu_k(X)$ is $L^k$. For $Y=A X$, it follows from the identity 
\begin{align}\label{e140}
(A_1  B_1) \otimes (A_2 B_2) = ( A_1  \otimes A_2 ) ( B_1 \otimes B_2 ),
\end{align}
where $A_1, A_2, B_1, B_2$ are matrices of suitable 
dimensions \cite[p. 408]{Lancaster:1985}, that 
\begin{align}\label{e13}
\mu_k(Y) = \underset{k \textrm{ times }}{\underbrace{ \left ( A  \otimes A  \otimes \cdots \otimes A  \right )}} \ \mu_k(X).   
\end{align}

Expressions for the cumulants of the observed process in a state-space model were 
developed by Swami and Mendal \cite{Swami:1990}. The process in this paper is 
a particular case for which simpler expressions hold. We provide an alternative 
derivation for our particular case in the Appendix.
Let $K_k(X)$ denote the vectorized $k$th central cumulant of any real random vector $X$ with finite mean. 
For $k=1,2,3$, $K_k(X)=\mu_k(X)$. For $k=4$ we have 
\begin{align}\label{e9}
K_4(X) &= \mu_4(X) - \mu_2(X) \otimes \mu_2(X) \nn 
&- \rvec \left \{ R_{X X } \otimes R_{X X } 
+ U_{L^2} \cdot (R_{X X } \otimes R_{X X }) \right \} 
\end{align}
where $R_{X X }$ is the covariance matrix of $X$ and $U_{L^2}$ is an $L^2 \times L^2$ 
permutation matrix defined in (\ref{a6}). 
We also have 
\begin{align}\label{e10}
K_4(Y) = (A \otimes A \otimes  A \otimes A) K_4(X). 
 \end{align}
When $R_{X  X }$ is a diagonal matrix, as is the case of interest in this paper, 
it follows that $U_{L^2} \cdot (R_{X X } \otimes R_{X X })$ is a symmetric matrix. 
We shall restrict our attention to cumulants of order $k \leq 4$ since estimation of higher order cumulants is not practical.

Assume next that the components $\{x_j\}$ of $X$ are 
independent. For this special case of interest, the central moments and central cumulants of $X$ 
are conveniently and concisely expressed in terms of the Khatri-Rao product defined by 
 \begin{align}\label{e11}
A \odot A := [a_1 \otimes a_1, a_2 \otimes a_2,\ldots,a_L \otimes a_L]. 
\end{align}
It is shown in the Appendix that
\begin{align}\label{e15}
 \mu_2(Y) &= (A \! \odot \!  A )  \col \!  \left (E \left \{\bar{x}_1^2 \right\},E \left \{\bar{x}_2^2 \right\}, \ldots, E \left \{\bar{x}_L^2 \right\} \right) \nn 
 \mu_3(Y) &= (A \! \odot  \! A   \odot \! A)  
 \col \! \left (E \left \{\bar{x}_1^3 \right \},E \left \{\bar{x}_2^3 \right\}, \ldots, E \left \{\bar{x}_L^3 \right\} \right ) ,
\end{align}
and 
\begin{align}\label{e16}
   K_4(Y) &=  (A \! \odot \! A  \! \odot \! A \! \odot \! A) \left ( \begin{array}{c}
 \kappa_4(\bar{x}_1, \bar{x}_1, \bar{x}_1, \bar{x}_1) \\ 
 \kappa_4(\bar{x}_2, \bar{x}_2, \bar{x}_2, \bar{x}_2) \\ 
 \vdots \\ 
 \kappa_4(\bar{x}_L, \bar{x}_L, \bar{x}_L, \bar{x}_L) 
 \end{array} \right ),   
\end{align}
 where 
  \begin{align}\label{e17}
  \kappa_4(\bar{x}_i, \bar{x}_j, \bar{x}_k, \bar{x}_l) = \left \{
  \begin{array}{cc}
  E\{\bar{x}_i^4\}  -  3 (E\{\bar{x}_i^2\})^2, & i \! = \! j \! = \! k \! = \! l \\ 
  0, & \textrm{otherwise.} 
  \end{array} 
  \right . 
   \end{align} 
 When $\{x_j\}$ are independent Poisson random 
 variables, all cumulants of $x_j$ equal its rate $E\{x_j\}=\lambda_j$, and we have from 
 $m(Y):= E\{Y\} = A E\{X\}$ and (\ref{e15})-(\ref{e17}) that, 
 \begin{align}\label{e7}
 \left (
 \begin{array}{c}
 A \\ 
 A \odot A \\ 
 A \odot A \odot A  \\ 
 A \odot A \odot A \odot A 
 \end{array} 
 \right ) \lambda = 
 \left (
 \begin{array}{c}
 m(Y) \\ 
 \mu_2(Y) \\ 
 \mu_3(Y)  \\ 
 K_4(Y)
 \end{array} 
 \right ). 
 \end{align}
 This is the key cumulant matching equation for $r \leq 4$. In \eqref{e7}, let $\cA_r$ 
 denote the matrix of stacked Khatri-Rao 
 products of $A$, and let $\eta_r(Y)$ denote the vector of stacked 
 cumulants. Then the rate vector $\lambda$ satisfies 
\begin{align}\label{e20}
 \cA_r \lambda = \eta_r(Y).   
 \end{align} 
 To estimate $\lambda$, 
 $\eta_r(Y)$ is replaced by a vector of empirical estimates $\hat{\eta}_r(Y)$ 
 and $\lambda$ is estimated from  
 \begin{align}\label{e139}
\boxed{ \cA_r \lambda = \hat{\eta}_r(Y).}   
 \end{align}
 In this paper we consider least squares estimation and 
 the minimum I-divergence iterative procedure (\ref{e77}). 
 The vector $\hat{\eta}_r(Y)$ comprises 
 minimum variance unbiased cumulant estimates given by the   
$K$-Statistics which we discuss in the next section \cite{Kendall:1969}.

 Vardi's second-order moment matching equation can be summarized as $\hat{\eta}_2(Y)=\cA_2 \lambda$. 
 With only two moments, the column rank of the matrix $\cA_2$ may be too low to provide an 
 accurate solution. Note that if the set of equations \eqref{e139} is consistent, 
 the theoretical cumulants are known, and 
 $r$ is sufficiently large so that $\cA_r$ has full column rank, 
 then $\lambda$ can be estimated 
 error free as the unique solution of \eqref{e139}. We demonstrate in Section \ref{num} for the 
 NSFnet \cite{Bonani:2016} that $r=3$ is sufficient to achieve full column rank, and that the 
 theoretical performance is approachable when the cumulants are 
 estimated from a sufficiently large number of realizations of $Y$.

\subsection{Error Analysis} \label{mainC}

In this section we assess the MSE in the least squares estimation  
of $\lambda$ satisfying (\ref{e20}) from the cumulant matching equation \eqref{e139}. 
We assume for this analysis that $r$ is sufficiently large so that 
the augmented matrix $\cA_r$ has full column rank.

$K$-Statistics of scalar processes were developed in 
\cite[p. 281]{Kendall:1969}. For vector processes and $k=1,2,3$, 
let $\hat{\mu}_k(Y)$ and $\hat{\mu}_k(X)$ denote the $K$-Statistics of $\mu_k(Y)$ and $\mu_k(X)$, 
respectively. Similarly, let $\hat{K}_4(Y)$ and $\hat{K}_4(X)$ denote the $K$-Statistics of 
$K_4(Y)$ and $K_4(X)$, respectively. Let $\{X_n, n=1,2, \ldots, N\}$ denote a sequence  
of independent identically distributed source-destination traffic flows defined similarly 
to $X$. Each $X_n$ comprised $L$ independent Poisson random variables with rate $E\{X_n\}=\lambda$. 
Let $Y_n=A X_n$. Define 
\begin{align}\label{e43}
\tilde{Y}_n := Y_n - \hat{m}(Y) 
\end{align}
where 
\begin{align}\label{e42}
\hat{m}(Y) :=\frac{1}{N}\sum_{n=1}^N Y_n,
\end{align}
and the empirical cumulant estimate for $k > 1$ as
\begin{align}\label{e84}
\tilde{\mu}_k(Y) := \frac{1}{N} \sum_{n=1}^N \underset{k \textrm{ times }}{\underbrace{
\tilde{Y}_n \otimes \tilde{Y}_n \otimes \cdots \otimes \tilde{Y}_n}} . 
\end{align}
Similar definitions follow for the $X$ process when $Y$ in (\ref{e43})-(\ref{e84}) is replaced by 
$X$. The $K$-Statistic for $k=1$ is given by 
\begin{align}\label{e83}
\hat{m}(Y) = \frac{1}{N}\sum_{n=1}^N A X_n = A \hat{m}(X),   
\end{align}
for $k=2$,
\begin{align}\label{e44}
\hat{\mu}_2(Y) &= \frac{N}{N-1} \tilde{\mu}_2(Y) \nn
&= (A \otimes A) \hat{\mu}_2(X), 
\end{align}
for $k=3$, 
\begin{align}\label{e45}
\hat{\mu}_3(Y) &= \frac{N^2}{(N-1)(N-2)} \tilde{\mu}_3(Y) \nn
&= (A \otimes A \otimes A) \hat{\mu}_3(X)   
\end{align}
and for $k=4$, 
\begin{align}\label{e46}
\hat{K}_4(Y) &:= \frac{N^2}{(N-1)(N-2)(N-3)} \left [ (N+1) \tilde{\mu}_4(Y) \right. \nn 
& \hspace{15ex} \left. - 3(N-1) \tilde{\mu}_2 (Y) 
\otimes \tilde{\mu}_2 (Y) \right ]\nn 
&= (A \otimes A \otimes A \otimes A) \hat{K}_4(X),  
\end{align}
where we have used (\ref{e140}). Similar relations are expected to hold for higher order cumulants. 
For $r \leq 3$ and sufficiently large $N$, the $K$-Statistic $\hat{\mu}_r(Y)$ and the empirical 
cumulant $\tilde{\mu}_r(Y)$ are essentially the same. 

Let $\hat{\eta}_r(X)$ and $\hat{\eta}_r(Y)$ denote vectors of stacked $K$-statistics of order 
smaller than or equal to $r$ corresponding to
$X$ and $Y$, respectively. The least squares estimate of $\lambda$ in \eqref{e139} is given by 
\begin{align}\label{e37}
\hat{\lambda}_r = (\cA_r^{*} \cA_r)^{-1} \cA_r^{*} \ \hat{\eta}_r (Y).   
\end{align}
Let  
\begin{align}\label{e48}
\epsilon_r (Y) = \hat{\eta}_r(Y) - \eta_r(Y)
\end{align}
denote the error vector in estimating $\eta_r(Y)$. From (\ref{e20}),  
\begin{align}\label{e36}
 \hat{\eta}_r(Y) = \cA_r \lambda + \epsilon_r (Y).
\end{align}
Substituting (\ref{e36}) into (\ref{e37}) yields,
\begin{align}\label{e38}
\hat{\lambda}_r &= (\cA_r^{*} \cA_r)^{-1} \cA_r^{*} [\cA_r \lambda + \epsilon_r (Y)] \nn
&= \lambda + (\cA_r^{*} \cA_r)^{-1} \cA_r^{*} \epsilon_r (Y).
\end{align}
Let $H_r=(\cA_r^{*} \cA_r)^{-1} \cA_r^{*}$. The rate estimation error is given by 
\begin{align}\label{e39}
\hat{\lambda}_r -  \lambda =  H_r \epsilon_r (Y), 
\end{align}
and the MSE is given by 
\begin{align}\label{e55}
\overline{\epsilon_r^2} (\lambda)  &= 
E\left \{ ||\hat{\lambda}_r -  \lambda||^2 \right \} \nn 
&= E\left \{ ||H_r \epsilon_r (Y)||^2 \right \} \nn
&\leq \rho_{\textrm{max}}(H_r^{*} H_r) E \{ ||\epsilon_r (Y)||^2\} 
\end{align}
where the inequality follows from the Rayleigh quotient theorem \footnote{
The Rayleigh quotient of a Hermitian matrix $H$ is defined as $u^{\ast} H u /u^{\ast} u$ where $u$ is a vector of suitable dimension.}  
\cite[Theorem 8.1.4]{Lancaster:1985},
and $\rho_{\textrm{max}}(H_r^{*} H_r)$ denotes the maximal eigenvalue of the Hermitian matrix $H_r^{*} H_r$.
From the theory of singular value decomposition, we recall that for any matrix $H_r$, 
\begin{align}\label{e128}
\rho_{\textrm{max}}(H_r H_r^{*}) = \rho_{\textrm{max}}(H_r^{*} H_r).
\end{align}
Applying this to (\ref{e55}) gives 
\begin{align}\label{e41}
\overline{\epsilon_r^2}(\lambda) & \leq  
\rho_{\textrm{max}} (((\cA_r^{*} \cA_r)^{-1} \cA_r^{*}) ((\cA_r^{*} \cA_r)^{-1} \cA_r^{*})^{*} ) E \{ ||\epsilon_r (Y)||^2\} \nn
&= \rho_{\textrm{max}}((\cA_r^{*} \cA_r)^{-1}) E \{ ||\epsilon_r (Y)||^2\} \nn
&=\frac{1}{\rho_{\textrm{min}}(\cA_r^{*} \cA_r)} E \{ ||\epsilon_r (Y)||^2\}.
\end{align}

Let $\cB_r$ denote a block diagonal 
matrix with the $k$th diagonal block being $\cA_k$. There are $r$ diagonal blocks
in total. Let  
\begin{align}\label{e40}
\epsilon_r (X) = \hat{\eta}_r(X) - \eta_r(X).  
\end{align}
From (\ref{e48}), (\ref{e83})-(\ref{e46}), and (\ref{e7}), 
\begin{align}\label{ee82}
\epsilon_r (Y)= \cB_r \epsilon_r (X) 
\end{align}
It follows from the Rayleigh quotient \cite[Theorem 8.1.4]{Lancaster:1985} that 
\begin{align}\label{e49}
||\epsilon_r (Y) ||^2 \leq ||\cB_r||^2 \cdot ||\epsilon_r (X)||^2
\end{align}
where $||\cB_r||$ equals the maximal singular value of $\cB_r$, and 
\begin{align}\label{e50}
||\cB_r||^2 =\rho_{\textrm{max}}(\cB_r^{*} \cB_r).
\end{align}
The $k$th diagonal block of $\cB_r^{*} \cB_r$ is given by 
\begin{align}\label{e51}
 (A \otimes A \otimes \cdots & \otimes A)^{*} (A \otimes A \otimes \cdots \otimes A) \nn
 &= A^{*} A \otimes A^{*} A \otimes \cdots \otimes A^{*} A  \nn
 & =: (A^{*} A )^{\otimes k} .
 \end{align} 
Since the eigenvalues of the Kronecker product of two matrices equal 
the Kronecker product of the vectors of eigenvalues of each matrix \cite[p. 412]{Lancaster:1985}, we conclude 
that the eigenvalues of $(A^{*} A )^{\otimes p}$ for any positive 
integer $p$ are all possible products of length $p$ of the eigenvalues 
of $A^{*}A$. In particular, since $A^{*}A$ is positive semi-definite, 
\begin{align}\label{e52}
 \rho_{\textrm{max}}((A^{*} A )^{\otimes p}) = [ \rho_{\textrm{max}}(A^{*} A )]^p . 
 \end{align} 
Furthermore, since $\cB_r^{*} \cB_r$ is block diagonal with blocks of increasing size, its 
eigenvalues are given by the union of the sets of eigenvalues associated with each block. Thus, 
\begin{align}\label{e53}
\rho_{\textrm{max}}(\cB_r^{*} \cB_r) = \max_{p: p \geq 1} [ \rho_{\textrm{max}}(A^{*} A )]^p.
\end{align}
When $\rho_{\textrm{max}}(A^{*} A ) > 1$, the maximum in (\ref{e53}) is achieved for 
$p=r$. Otherwise, when $\rho_{\textrm{max}}(A^{*} A ) < 1$, the maximum in (\ref{e53}) 
is achieved for $p=1$. Hence, from (\ref{e49}), 
\begin{align}\label{e54}
E\{||\epsilon_r (Y)||^2\} &\leq \max \left \{ \rho_{\textrm{max}}(A^{*} A ), 
[\rho_{\textrm{max}}(A^{*} A )]^r \right \} \nonumber \\
 & \hspace{20ex} 
 \cdot E \left \{||\epsilon_r (X)||^2 \right \}.
\end{align}
From (\ref{e41}) and (\ref{e54}) we obtain, 
\begin{align}\label{e57}
\overline{\epsilon_r^2}(\lambda) & \leq  
\frac{\max \left \{ \rho_{\textrm{max}}(A^{*} A ), 
[\rho_{\textrm{max}}(A^{*} A )]^r \right \}}{\rho_{\textrm{min}}(\cA_r^{*} \cA_r)} E \{ ||\epsilon_r (X)||^2\}.
\end{align}

Evaluating the MSE $E \{ ||\epsilon_r (X)||^2\}$ is quite involved. For $r=2,3$, and sufficiently large 
$N$, the cross terms in $\hat{\mu}_r(X)$ will be approximately zero, and hence    
\begin{align}\label{e121}
E \{ ||\epsilon_r (X)||^2\} \approx \sum_{l=1}^L E \{ ||\epsilon_r (x_l)||^2\}   
\end{align}
where $E \{ ||\epsilon_r (x_l)||^2\}$ is the MSE associated with the $K$-statistics   
of the $l$th component of $X$. It is also the variance of $\hat{\mu}_r(x_l)$. 
Since the $\{X_n\}$ vectors are assumed independent, we can infer $E \{ ||\epsilon_r (x_l)||^2\}$ 
from the variance of the $K$-statistics of an IID sequence of scalar random variables 
given by \cite[p. 291]{Kendall:1969}, 
\begin{align}\label{b20}
\var & \left( \hat{\mu}_2(x_l) \right) = \frac{1}{N} K_4(x_l) + \frac{1}{N-1} 2 \mu_2^2(x_l) \nn
\var & \left( \hat{\mu}_3(x_l) \right) = \frac{1}{N} K_6(x_l) + \frac{9}{N-1} K_4 (x_l) \mu_2 (x_l) 
  \nn
 & \hspace{2ex} + \frac{9}{N-1} \mu_3^2(x_l) + \frac{6 N}{(N-1)(N-2)} \mu_2^3(x_l) . 
\end{align}
When the IID random variables are Poisson with $E\{X_l\}=\lambda_l$, all cumulants equal $\lambda_l$, and hence, 
\begin{align}\label{b21}
E \{ ||\epsilon_2 (x_l)||^2\} &= \var \left( \hat{\mu}_2(x_l) \right) = \frac{1}{N} \lambda_l + \frac{1}{N-1} 2 \lambda_l^2 \nn
&\approx \frac{1}{N} (\lambda_l + 2 \lambda_l^2). 
\end{align} 
\begin{align}\label{b22}
E & \{ ||\epsilon_3 (x_l)||^2\} = \var \left( \hat{\mu}_3(x_l) \right) = \frac{1}{N} \lambda_l + \frac{9}{N-1} 2 \lambda_l^2 \nn
& ~~~ + \frac{6N}{(N-1)(N-2)} \lambda_l^3 
\approx  \frac{1}{N} (\lambda_l + 18 \lambda_l^2 + 6 \lambda_l^3). 
\end{align}
The upper bound on $\overline{\epsilon_r^2}(\lambda)$ for 
$r=2,3$, follows from (\ref{e57}), (\ref{e121}), (\ref{b21}) and (\ref{b22}). 

The bound (\ref{e57}) is loose when the condition number of the matrix $\cA_r^* \cA_r$ 
is large. It does have qualitative value as it shows that estimation of moments becomes 
increasingly harder as the order increases. Since usually $\lambda_l >1$ in network tomography, 
the error in estimating $\mu_3(x_l)$ is much larger than that in estimating $\mu_2(x_l)$.

\subsection{Implementation} \label{implementation}

In this section we address several aspects related to the implementation of the 
cumulant matching approach. The Khatri-Rao product can be expressed in terms of the rows
of $A$ instead of its columns as in (\ref{e11}).  Let $\alpha_i$ denote the $i$th row
of $A$, $i = 1, \ldots, M$.  Then $A = \mbox{stack} \{  \alpha_i : i = 1, \ldots, M \}$
where $\mbox{stack}$ refers to the stacking of row vectors of equal length to form a matrix.  
We have 
\begin{align}
A \odot A &= \mbox{stack} \left \{ \alpha_i \circ \alpha_j; \ i,j \! = \! 1, \ldots, M \right \} \nn
A \odot A \! \odot \! A &= \mbox{stack} \left \{ \alpha_i \circ \alpha_j \circ \alpha_k; 
\ i,j,k \! = \! 1, \ldots, M \right \},
\end{align}
etc., where $\circ$ denotes the Schur-Hadamard product and lexicographic ordering of the 
indices $(i,j, k)$ is assumed. Using this formulation, it is easy to see that $\cA_r$ 
contains duplicate rows. For example, $\alpha_i$ in $A$ and $\alpha_i \circ \alpha_i$ in $A \odot A$ are 
duplicates since the elements of $A$ belong to $\{0,1\}$. Additionally, $\cA_r$ may contain 
null rows as is easy to see from (\ref{e11}). Thus, equations in (\ref{e7}) corresponding to 
duplicate and null rows in $\cA_r$ can be removed. We shall 
always opt to removing duplicate rows that correspond to higher-order cumulants 
rather than to low order cumulants since higher-order cumulants are harder to estimate.  
To simplify notation we shall henceforth assume that $\cA_r$ and the right hand side (RHS) vector 
$\eta_r(Y)$ of (\ref{e7}) are given in their reduced form. Similarly, we shall consider (\ref{e20}) 
and (\ref{e139}) as given in their reduced form. 

It is interesting to compare the number of rows in the original and reduced $\cA_r$. 
The number of rows in the original $\cA_r$ is given by 
\begin{align}\label{e102}
 \bar{n}_r(M) = M + M^2 + M^3 + \ldots + M^r =  \frac{M (M^r \! - \! 1)}{M -1} .
\end{align}
The maximum number of distinct rows in the reduced $\cA_r$ is counted as 
the sum of the maximum number of distinct rows contributed individually by $A$, 
$A \circ A$, $A \circ A \circ A$, etc., with the last contribution 
from the $r$-fold Khatri-Rao product of $A$ with itself. The 
contribution from the $i$th term is given by ${M \choose i}$, 
$i=1,\ldots,r$, which represents the number of unordered combinations 
of $i$ rows chosen without replacement from the given $M$ rows. 
Thus, the maximum number of distinct rows in the reduced $\cA_r$ is given by 
\begin{align}\label{e101}
n_r(M) = {M \choose 1} \! + \! {M \choose 2} \! + \! {M \choose 3} + \ldots + {M \choose r} .
\end{align}
For example, consider a network with a   
$4 \times 15$ routing matrix $A$ whose columns comprise all lexicographically ordered  
non-zero binary $4$-tuples. Here, $M=4$, and $\bar{n}_r(M)$ 
and $n_r(M)$ are shown in Table \ref{t2}. Note that for this example, the number of reduced 
equations coincides with the row rank of $\cA_r$ for each $r$. Furthermore, full column rank 
is achieved only when $r=4$. 

\begin{table}
\centering
\begin{tabular}{|c|c|c|c|c|c|} 
\hline 
$r $ & 1 & 2 & 3 & 4 \\ 
\hline 
$\bar{n}_r(M)$ & $4$ & $20$  & $64$  & $320$  \\  
\hline 
$n_r(M)$ & $4$ & $10$  & $14$  & $15$  \\  
\hline 
row rank of $\cA_r$ & $4$ & $10$  & $14$  & $15$  \\  
\hline 
\end{tabular} 
\caption{Number of unreduced ($\bar{n}_r(M)$) and reduced ($n_r(M)$) cumulant matching equations, 
and row rank of $\cA_r$,  for the $4 \times 15$ routing matrix of all non-zero binary $4$-tuples.}
 \label{t2}
 \end{table}

We study two estimates of $\lambda$ that approximate the cumulant matching equation \eqref{e139}. 
The first estimate follows from the iteration \eqref{e77} which is initialized by a constant 
vector. The second estimate follows from least squares. 
Specifically, we use the unique Tikhonov regularized least squares solution for the inconsistent 
set of equations~\eqref{e139}, when $\mathcal{A}_r$ is not necessarily full column rank. This estimator 
is given by~\cite[p. 51]{Kailath:2000}
\begin{align}\label{e90}
 \hat{\lambda}_r  = (\cA_r^{*} \cA_r+\gamma I) ^{-1} \cA_r^{*} \ \hat{\eta}_r (Y)
 \end{align}
for some $\gamma > 0$. Note that the regularized estimator applies to a skinny as well as 
a fat matrix $\cA_r$.

To mitigate the effects of the error introduced by the empirical cumulant estimates, while 
allowing the estimator of $\lambda$ 
to benefit from the higher order cumulants, the relative weights 
of the third and fourth order cumulants compared to the first and second order moments, 
can be reduced. 
This can be done by multiplying all equations in (\ref{e90}) with rows originating from 
$\cA_3$ by some $0 < \epsilon_3 < 1$, 
and all equations with rows originating from $\cA_4$ by some $0 < \epsilon_4 < 1$. 
This regularization approach was advocated by Vardi \cite{Vardi:1996} in the context of 
his second-order moment matching approach. Following this approach, let the reduced and 
$\epsilon$-weighted matrix $\cA_r$ be denoted by $\cA_{r,\epsilon}$, and let the reduced  
$\epsilon$-weighted vector $\hat{\eta}_{r,\epsilon}(Y)$ be denoted by 
$\hat{\eta}_{r,\epsilon}(Y)$. Then, from (\ref{e90}), the rate vector $\lambda$ is estimated from 
\begin{align}\label{e92}
 \hat{\lambda}_{r,\epsilon}  = (\cA_{r,\epsilon}^{*} \cA_{r,\epsilon}+\gamma I) ^{-1} \cA_{r,\epsilon}^{*} \ 
 \hat{\eta}_{r,\epsilon} (Y).
 \end{align}
 
Note that the estimator \eqref{e92} is not guaranteed to be non-negative. 
A non-negative estimate of $\lambda$ can be obtained by using non-negative 
least squares optimization \cite{Lawson:1974}. 
In our numeric examples we have arbitrarily substituted negative estimates with the value of $.005$. This 
approach resulted in substantially lower MSE compared to using the constrained optimization 
algorithm of \cite[p. 161]{Lawson:1974}. The performance of the algorithm remained essentially 
the same when other small values (e.g., $.1$ or $.2$) were used since occurrences of 
negative estimates are infrequent at our working point. See Table~\ref{tbl:LS-lambda_est_Neg}.
 
 \subsection{Complexity} \label{complex}
 
The computational effort in the cumulant matching approach consists of 
the effort to construct and solve the set of equations \eqref{e139}. The 
number of equations in this set is $n_r(M)$. 
Construction of the left hand side of \eqref{e139} requires $(M^2+M^3+\ldots +M^r)L$ operations. 
Construction of the right hand side of \eqref{e139} requires $(M^2+M^3+\ldots +M^r)N$ operations 
where $N$ is the number vectors used to estimate each cumulant. Solving the equations 
requires effort that depends only on $n_r(M)$, $M$ and the number of iterations in 
\eqref{e77}. The combined effort is dominated 
by $(M^2+M^3+\ldots +M^r)N$ since $N$ must be large to produce meaningful cumulant 
estimates. Thus the computational effort of the cumulant matching approach is 
approximately linear in $N$ when $N$ is large which is always the case.

\section{Random Routing} \label{rand}

In a typical network, there are multiple paths connecting every source node with every destination 
node. When the network operates under a deterministic routing regime, a single fixed path is 
used for every source-destination pair. When the network operates under a random routing regime, a 
path from the source to destination nodes is selected according to some probability law. Vardi 
attributed Markovian routing for traffic on each source-destination pair. The accessible nodes and 
links for the given pair are represented by the states and transition probabilities of the 
Markov chain, respectively. The transition probabilities of the Markov chain 
determine the probability of each path with the same source-destination address. 

Tebaldi and West argued that random routing can be viewed as deterministic routing in a super-network in 
which all possible paths for each source-destination address are listed \cite{Tebaldi:1998}. 
This approach results in an expanded zero-one routing matrix with each column in the original routing 
matrix replaced by multiple columns representing the feasible paths for the source-destination pair. 
The Poisson traffic flow on a source-destination pair is thinned into multiple Poisson traffic flows 
with the multinomial thinning probabilities being the probabilities of the paths with the 
same source-destination address. 
Thus, the super network and the original network operate under similar statistical 
(Poisson) models. The thinned Poisson rates can now be estimated as was originally done, e.g., using  
cumulant matching, and each source-destination rate estimate can be obtained from the sum of the thinned rate 
estimates in that source-destination pair.

\section{Numerical Example} \label{num}

\begin{figure*}
\centerline{
\includegraphics[scale=.37, clip=true]{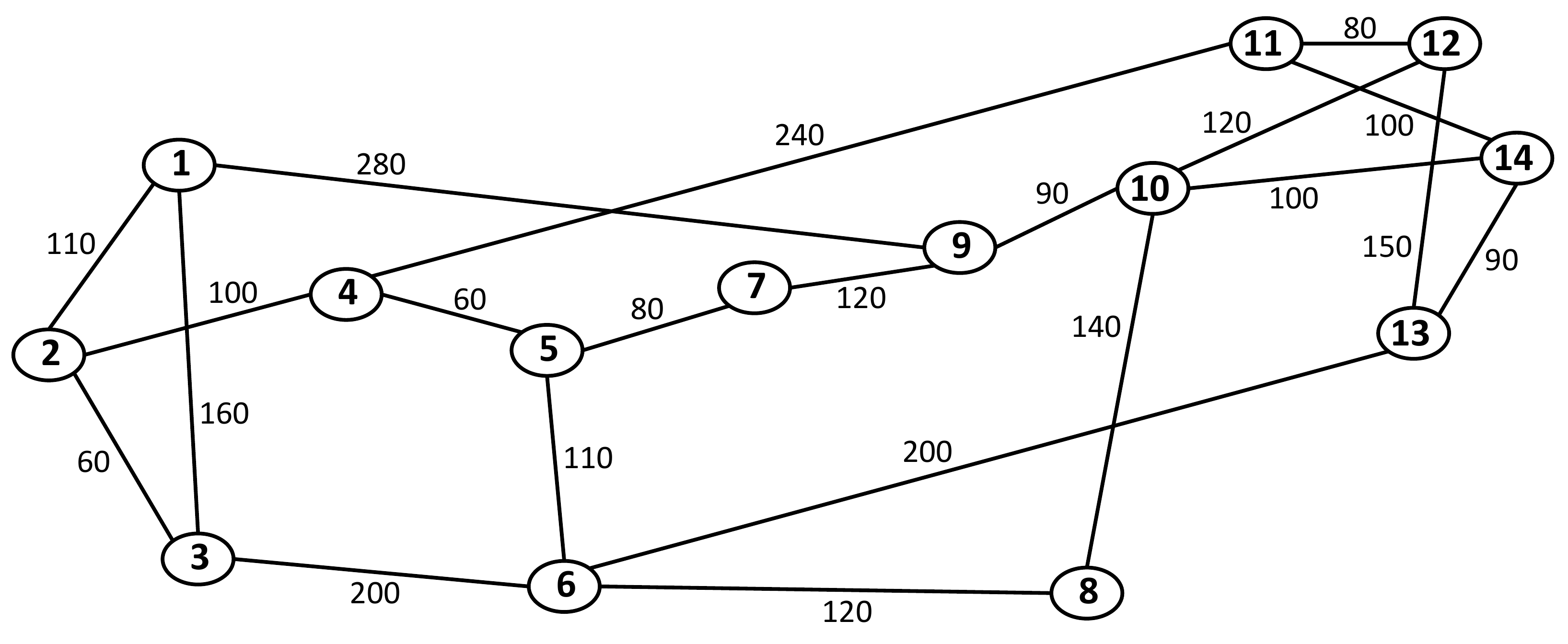}
}
\caption{NSFnet topology with link weights as in \cite[Fig. 4]{Bonani:2016}.}
\label{fig:nsfnet}
\end{figure*}

In this section we demonstrate the performance of our approach and the gain 
realized by using higher-order empirical cumulants\footnote{Our experiments 
were run on ARGO, a research computing cluster
provided by the Office of Research Computing at George Mason University,
VA. (URL: http://orc.gmu.edu).}. We study the NSFnet \cite{Bonani:2016}
whose topology is shown in Fig. \ref{fig:nsfnet}. The network consists of 14 nodes 
and 21 bidirectional links. Hence, it contains $L=14 \cdot 13 /2 =91$ source-destination pairs. 
This size network may represent a private network, a transportation network or a subset of interest of 
a large-scale network.
The link weights in Fig. \ref{fig:nsfnet} 
are exclusively used to determine $k \geq 1$ shortest paths for each source-destination pair. 
Otherwise, they play no role in the traffic rate estimation problem.
To determine the $k$ shortest paths between a given
source-destination pair, we used the shortest simple paths
function from the NetworkX Python library, which is based on
the algorithm of Yen \cite{Yen:1971}.
When $k=1$, the number of source-destination paths equals the number of source-destination pairs and the  
routing matrix $A$ is a $21 \times 91$ matrix. The augmented routing matrix 
$\cA_r$ achieves full column rank when $r=2$.

With $k \geq 2$, we can assign multiple paths to each source-destination pair and treat 
them as distinguishable new source-destination pairs. The routing matrix $A$ thus 
becomes fatter and using higher-order empirical cumulants 
becomes beneficial. For example, when $k=2$, we have $L=182$ source-destination paths, 
the column rank of $\cA_2$ is $162$ and $\cA_3$ has full column rank.
Thus, using this example we focus on third-order cumulant matching. 

The network with $k=2$ and a $21 \times 182$ routing matrix $A$ could also 
be seen as a super-network in the Tebaldi-West sense \cite{Tebaldi:1998}  for a network 
with $L=91$ source-destination pairs operating under a random routing regime with two 
possible paths per each source-destination pair. The accuracy of the rate estimation 
for the random routing network is determined by the accuracy of the rate estimation in 
the deterministic routing super-network. Thus, it suffices to focus on rate estimation in the 
deterministic routing network with the $21 \times 182$ routing matrix $A$. 

The arrival rates $\{\lambda_j\}$ in our experiment were generated randomly from the interval
$[0,4]$. In each of $T=500$ simulation runs, a rate vector $\lambda$ was generated, and  
was subsequently used in 
generating $N$ statistically independent identically distributed Poisson vectors $\{X_n\}$ which 
were transformed into the vectors $\{Y_n=AX_n\}$ using the assumed known routing matrix $A$. The $N$ 
statistically independent identically distributed Poisson vectors $\{Y_n\}$ were used to generate 
the empirical cumulants using \eqref{e42} when $r=1$ and \eqref{e84} when $r=2,3$. 
We  experimented with $N$ in the range of 
$10\,000$ to $500\,000$. The MSE in estimating the cumulants for the NSFnet with $M=21$ are given 
in Table \ref{tbl:Cum-Est}. Clearly, the MSE decreases monotonically with $N$ for 
$r=1,2,3$.

The empirical cumulants were used to estimate the rate vector in the current run. 
For the cumulants regularization we have used $\epsilon_2=1$ and $\epsilon_3=.01$.
Let $\lambda_t(i)$ and $\hat{\lambda}_t(i)$ denote, respectively, 
the $i$th component of $\lambda$ and its estimate at the $t$th run where $i=1,\ldots,L$ and 
$t =1,\ldots , T$. For each estimate we evaluated the \textit{normalized MSE} defined by 
\begin{align}\label{e105}
\xi_i^2 = \frac{\frac{1}{T} \sum_{t=1}^T (\lambda_t(i)-\hat{\lambda}_t(i))^2}
{\frac{1}{T} \sum_{t=1}^T (\lambda_t(i))^2} 
\end{align}
and the \textit{averaged normalized MSE}  defined by 
\begin{align}\label{e106}
\overline{\xi^2} = \frac{1}{L} \sum_{i=1}^L \xi_i^2. 
\end{align}
The MSE in estimating $\lambda_i$ is approximately 
$\xi_i^2 \cdot E\{\lambda^2(i)\}$ when $T$ is sufficiently large. 

Two rate estimators were used, the iterative estimator \eqref{e77} and the least squares estimator \eqref{e92}. 
The iteration was initialized uniformly with all rates set to $.1$. It was terminated 
after $300$ iterations. The least squares regularization factor was set to $\gamma=.0005$.

  \begin{table}
\begin{center}
\begin{tabular}{|r|rrr|}
 \cline{1-4} 								
$N$	       & $r=1$	&   $r=2$  & 	$r=3$    \\  \hline
$10\,000$  & 0.0048 & 	0.2578 & 	14.9097  \\
$20\,000$  & 0.0025 & 	0.1285 & 	7.4569	  \\
$50\,000$  & 0.0010	&   0.0515 & 	2.9978	   \\   
$100\,000$ & 0.0005 & 	0.0257 & 	1.4989   \\
$200\,000$ & 0.0002 & 	0.0131 & 	0.7429	   \\
$500\,000$ & 0.0001 & 	0.0052 &    0.2993   \\  \hline	
\end{tabular}
\end{center}
\caption{MSE in cumulant estimation for the NSFnet   
with $M=21$ links.}
 \label{tbl:Cum-Est}
 \end{table}	
 
\begin{table}
\begin{center}
\begin{tabular}{|r|rrr|}
 \cline{1-4} 								
$N$	       & $r=1$	&   $r=2$  & 	$r=3$    \\  \hline
$10\,000$  & 0.2292 & 	0.0993 & 	0.1008  \\
$20\,000$  & 0.2292 & 	0.0701 & 	0.0694	  \\
$50\,000$  & 0.2292	&   0.0502 & 	0.0467	   \\   
$100\,000$ & 0.2292 & 	0.0434 & 	0.0381   \\
$200\,000$ & 0.2292 & 	0.0399 & 	0.0335	   \\
$500\,000$ & 0.2292 & 	0.0372 &    0.0298   \\  \hline	
Theoretical& 	0.2292 & 	0.0353	 & 0.0015  \\  \hline
Rank$(\cA_r)$	& 21 &	162 &	182	 \\  \hline	
\end{tabular}
\end{center}
\caption{Averaged normalized MSE $\overline{\xi^2}$ 
in $L=182$ source-destination path rate estimation of   
          the NSFnet using the Iteration~\eqref{e77}.}
 \label{tbl:MID-lambda_est}
 \end{table}	
 		
\begin{table}
\begin{center}
\begin{tabular}{|r|rrr|}
 \cline{1-4} 								
$N$	       & $r=1$	&   $r=2$  & 	$r=3$    \\  \hline
$10\,000$  & 0.3037 & 	0.0951 & 	0.0950  \\
$20\,000$  & 0.3037 & 	0.0637 & 	0.0625	  \\
$50\,000$  & 0.3037	&   0.0430 & 	0.0407	   \\   
$100\,000$ & 0.3037 & 	0.0358 & 	0.0329   \\
$200\,000$ & 0.3037 & 	0.0321 & 	0.0288	   \\
$500\,000$ & 0.3037 & 	0.0293 &    0.0257   \\  \hline	
Theoretical& 	0.3037 & 	0.0275	 & 0.0000  \\  \hline
Rank$(\cA_r)$	& 21 &	162 &	182	 \\  \hline	
\end{tabular}
\end{center}
\caption{Averaged normalized MSE $\overline{\xi^2}$ 
in $L=182$ source-destination path rate estimation of   
          the NSFnet using the Least Squares Estimator~\eqref{e92}.}
 \label{tbl:LS-lambda_est}
 \end{table}

  \begin{table}
\begin{center}
\begin{tabular}{|r|rrr|}
 \cline{2-4} 	
 \hline	
$N$	    & $r=1$	 & $r=2$  & 	$r=3$  	 \\  \hline   
$10\,000$ 	& 0.1890	&4.1747	& 4.2275  \\
$20\,000$	& 0.1868	&2.9813	& 3.0626  \\
$50\,000$	& 0.1890	&1.9879	& 1.9901  \\
$100\,000$	& 0.1890	& 1.6385	& 1.5780  \\
$200\,000$	& 0.1890	& 1.3747	& 1.3242  \\
$500\,000$	& 0.1890	& 1.1956	& 1.1165  \\  \hline 
Theoretical	& 0.1890	& 0.9692	& 0.0022 \\  \hline 
\end{tabular}
\end{center}			
\caption{Percent of Negative rate estimates in Table \ref{tbl:LS-lambda_est}.}
\label{tbl:LS-lambda_est_Neg}
\end{table}

The performance of the iterative estimator \eqref{e77} is demonstrated in
Table~\ref{tbl:MID-lambda_est}. The table shows the averaged normalized MSE $\overline{\xi^2}$ 
in estimating the path rates using $r= 1, 2, 3$ empirical cumulants estimated 
from $N$ vectors $\{Y_n\}$ with $N$ ranging from $N=10\,000$ to $N=500\,000$. 
The averaged normalized MSE $\overline{\xi^2}$ in estimating the path rates 
using $r= 1, 2, 3$ theoretical cumulants is shown as well. 
As can be observed in the table, $\overline{\xi^2}$ monotonically decreases with  
$N$ for $r=2,3$. For a given $N$, the performance using $r=3$ cumulants is  
better than using $r=2$ moments, except for $N=10\,000$ where the difference is only
$.0005$. Similar behavior can be seen in Table~\ref{tbl:LS-lambda_est} when  the 
least squares estimator \eqref{e92} is used for the 
 path rate estimation. The percentage 
of negative rate estimates observed with the least squares estimator is shown in 
Table \ref{tbl:LS-lambda_est_Neg}. This percentage is negligible for all values of $r$ and $N$ 
and does not exceed $5\%$. Comparing the two estimators \eqref{e77} and \eqref{e92}, 
the least squares estimator appears superior as it consistently provides lower 
$\overline{\xi^2}$ at the expense of  negligible percentage of negative rate 
estimates.  
 
To assess the effectiveness of the third-order cumulant matching approach and compare it 
with Vardi's second-order moment matching approach, we contrast the performance of the 
former approach with the \textit{best} performance of the latter approach as obtained by using the 
theoretical rather than the empirical moments. Let $\overline{\xi^2}(r,N)$ denote the averaged normalized 
MSE from \eqref{e106} obtained when $\lambda$ is estimated using $r$ empirical cumulants that were estimated from $N$ 
realizations of the vector $Y$. Let 
$\overline{\xi^2}(r,\infty)$ denote the averaged normalized MSE $\overline{\xi^2}$ from  \eqref{e106} obtained by using $r$ theoretical   
cumulants, or effectively by using $N=\infty$. Table \ref{tbl:VardiComp} shows $\overline{\xi^2}(2,N)/\overline{\xi^2}(2,\infty)$ 
and $\overline{\xi^2}(3,N)/\overline{\xi^2}(2,\infty)$ for $N=200\,000$ 
and the estimators \eqref{e77} and \eqref{e92} discussed  in Tables \ref{tbl:MID-lambda_est}--\ref{tbl:LS-lambda_est}. 

\begin{table}
\begin{center}
\begin{tabular}{|c|r|r|}
 \cline{1-3} 	
 	    & $\hat{\lambda}$ [Eq. \eqref{e77}]	 & $\hat{\lambda}$ [Eq. \eqref{e92}]  \\  \hline   
${\overline{\xi^2}(2,N)}/{\overline{\xi^2}(2,\infty)}$  	&	1.1303	& 1.1673   \\  
${\overline{\xi^2}(3,N)}/{\overline{\xi^2}(2,\infty)}$		&0.9490 & 1.0473  \\  \hline 
\end{tabular}
\end{center}			
\caption{Relative averaged normalized MSE $\overline{\xi^2}$ in second and third order cumulant matching with $N=200\,000$, to minimum 
second-order averaged normalized MSE.}
\label{tbl:VardiComp}
\end{table}
 
 Table \ref{tbl:VardiComp} shows the relative averaged normalized MSE in second and 
 third-order cumulant matching compared to the minimum averaged normalized MSE in 
 theoretical second-order moment matching as obtained by using the true second-order 
 moments.  The reduction in the relative averaged normalized MSE obtained when 
 the estimator \eqref{e77} was used was from $1.1303$ to $0.9490$, and from 
  $1.1673$ to $1.0473$ when the estimator \eqref{e92} was used. The third-order 
  cumulants in this experiment were estimated using $N=200\,000$.  
 This represents a reduction of about $12\%-18\%$.

\begin{figure}
\centerline{
\includegraphics[scale=.32, clip=true]{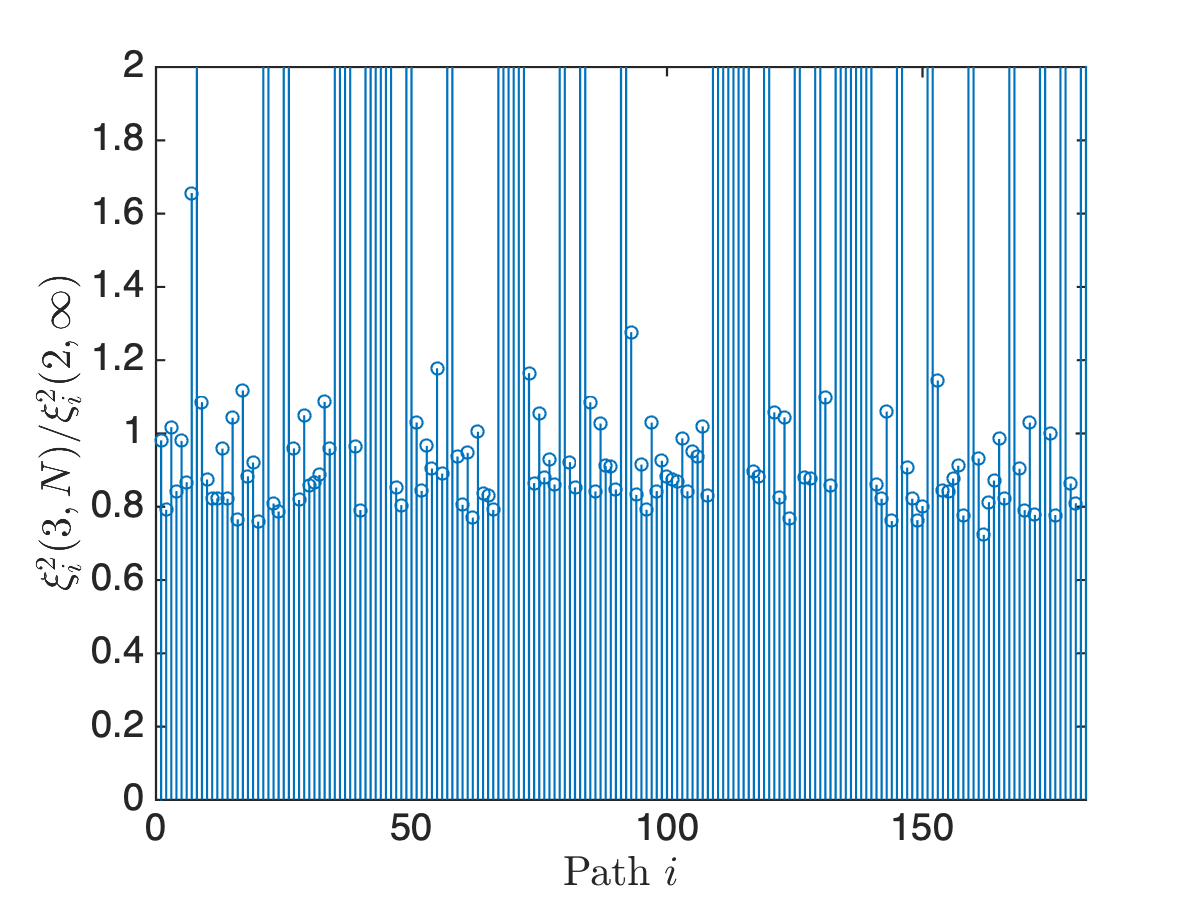}
}
\caption{Individual rate estimation performance as measured by the ratio $\xi_i^2(3,N)/\xi_i^2(2,\infty)$ 
for $i=1,\ldots, 182$. The rate vector $\lambda$ was estimated using the Iteration \eqref{e77} and $N=200\,000$. }
\label{fig:MID-Irate2_comp}
\end{figure}

To show the accuracy in estimating individual path rates, 
let $\xi_i^2(r,N)$ denote the normalized MSE $\xi_i^2$ from \eqref{e105} when 
estimating the $i$th component of $\lambda$ using $r$ empirical cumulants estimated from $N$ 
realizations of $Y$.  Figure  \ref{fig:MID-Irate2_comp}  
shows $\xi_i^2(3,N)/\xi_i^2(2,\infty)$ for $i=1,\ldots,182$ when estimation was done 
using the Iteration \eqref{e77} and $N=200\,000$.
Excluding outliers, the figure  shows that $\xi_i^2(3,N)$ equals about $.8-1.1$ of $\xi_i^2(2,\infty)$. 
The outliers correspond to rate components for which $\xi_i^2(2,\infty)$ is very small so that the ratio  
$\xi_i^2(3,N)/\xi_i^2(2,\infty)$ is rather large.

\section{Concluding Remarks} \label{comm}

We have developed a framework for high-order cumulant matching approach for estimating 
the rates of source-destination Poisson traffic flows from link traffic flows. 
The approach is equally applicable to networks operating under deterministic as well as 
random routing strategies. Under independent Poisson source-destination traffic flows, the 
approach boils down to a set of linear equations relating empirical cumulants of the link 
measurements to the rate vector $\lambda$ via a matrix involving Khatri-Rao products of 
the routing matrix. We studied an iterative minimum I-divergence approach and least 
squares estimation of the rate vector from the cumulant matching equations. We have 
established an upper bound on the MSE in least squares 
estimation of $\lambda$. The bound is useful for full column rank matrices $\cA_r$ with small 
condition number. We demonstrated the performance of the cumulant matching approach 
and compared it with Vardi's \textit{best} second-order moment matching on the NSFnet. 
We demonstrated that supplementing Vardi's approach with the third-order 
empirical cumulant reduces its \textit{minimum} averaged normalized MSE in rate estimation by $13\%-17\%$. 
The minimum averaged normalized MSE 
is obtained when the theoretical rather than the empirical moments are used in Vardi's second-order approach.
The computational complexity of the approach is linear in the number $N$ of link traffic flows 
used to estimate the cumulants.

\appendix

\setcounter{enumiv}{1}
\renewcommand{\theequation}{\Alph{enumiv}.\arabic{equation}}

\section{Cumulants of Linear Maps} \label{appA}

Let $e_i$ be an $L \times 1$ vector with a $1$ in the $i$th component and 
zeros elsewhere. The vector $\bar{X}$ can be written as  
$\bar{X}=\sum_{i=1}^L \bar{x}_i e_i$. 
Substituting in (\ref{b15}), and applying the distributive property of the Kronecker product, we obtain,  
\begin{align}\label{c7}
\mu_4(X) &= \sum_{i,j,k,l} E \left \{\bar{x}_i \bar{x}_j \bar{x}_k \bar{x}_l \right\} (e_i \otimes e_j \otimes e_k \otimes e_l).  
\end{align}
Similarly, the fourth order cumulant of $X$ is given by 
\begin{align}\label{c50}
K_4(X) &= \sum_{i,j,k,l} \kappa_4(\bar{x}_i, \bar{x}_j, \bar{x}_k, \bar{x}_l) (e_i \otimes e_j \otimes e_k \otimes e_l)  
\end{align}
where 
\begin{align}\label{c4}
\kappa_4 & (\bar{x}_i, \bar{x}_j, \bar{x}_k, \bar{x}_l) 
=E\{\bar{x}_i \bar{x}_j \bar{x}_k \bar{x}_l\} - E\{\bar{x}_i \bar{x}_j\} E\{ \bar{x}_k \bar{x}_l\}  \nn 
&  -E\{\bar{x}_i \bar{x}_k\} E\{ \bar{x}_j \bar{x}_l\} -E\{\bar{x}_i \bar{x}_l\} E\{ \bar{x}_j \bar{x}_k\}.    
\end{align}
Substituting (\ref{c4}) into (\ref{c50}) we obtain  
\begin{align}\label{c8}
K_4(X) 
& \overset{(i)}{=} \sum_{i,j,k,l} E\{\bar{x}_i \bar{x}_j \bar{x}_k \bar{x}_l\} (e_i \otimes e_j \otimes e_k \otimes e_l)  \nn 
& \overset{(ii)}{-} \sum_{i,j,k,l} E\{\bar{x}_i \bar{x}_j\} E\{ \bar{x}_k \bar{x}_l\} (e_i \otimes e_j \otimes e_k \otimes e_l) \nn 
& \overset{(iii)}{-}  \sum_{i,j,k,l} E\{\bar{x}_i \bar{x}_k\} E\{ \bar{x}_j \bar{x}_l\} (e_i \otimes e_j \otimes e_k \otimes e_l)  \nn 
& \overset{(iv)}{-}  \sum_{i,j,k,l} E\{\bar{x}_i \bar{x}_l\} E\{ \bar{x}_j \bar{x}_k\} (e_i \otimes e_j \otimes e_k \otimes e_l)  .
\end{align}
Row $(i)$ coincides with (\ref{c7}) and hence it equals $\mu_4(X)$.  
The expression in row $(ii)$ can be written as 
 \begin{align}\label{c9}
 &\sum_{i,j} E\{\bar{x}_i \bar{x}_j \} (e_i \otimes e_j)  \otimes \sum_{k,l} E \{\bar{x}_k \bar{x}_l\} (e_k \otimes e_l) \nn
 &\hspace{20ex} =  \mu_2(X) \otimes \mu_2(X). 
 \end{align}  
 For row $(iii)$, 
 \begin{align}\label{c12}
 (e_i \otimes e_j \otimes e_k \otimes e_l)  & = \rvec\{ (e_k \otimes e_l) (e_i \otimes e_j )^{*} \} \nn
 &= \rvec\{ (e_k \otimes e_l) (e_i^{*} \otimes e_j^{*} ) \} \nn
 &= \rvec\{(e_k e_i^{*}) \otimes (e_l e_j^{*}) \} ,  
 \end{align}
 and hence, 
\begin{align}\label{c11}
\sum_{i,j,k,l} E\{\bar{x}_i & \bar{x}_k\} E\{ \bar{x}_j \bar{x}_l\} (e_i \otimes e_j \otimes e_k \otimes e_l)  \nn 
 & = \rvec  \sum_{i,j,k,l} E\{\bar{x}_i \bar{x}_k\} E\{ \bar{x}_j \bar{x}_l\} (e_k e_i^{*} \otimes e_l e_j^{*})   \nn
 & = \rvec \left \{ R_{X X } \otimes R_{X X } \right \}.
\end{align}
For row $(iv)$ we utilize the $L^2 \times L^2$ permutation matrix 
\begin{align}\label{a6}
 U_{L^2} = \sum_{i=1}^{L} \sum_{j=1}^L (e_i e_j^{*}) \otimes (e_j e_i^{*}) 
 \end{align}
to obtain, 
\begin{align}\label{c13}
(e_i \otimes e_j \otimes e_k \otimes e_l) &=  (e_i \otimes e_j) \otimes \left ( U_{L^2} \cdot (e_l \otimes e_k) \right ) \nn 
&=  \rvec \left \{U_{L^2} \cdot (e_l \otimes e_k) \cdot (e_i \otimes e_j)^{*} \right\}\nn 
&=  \rvec \left \{U_{L^2} \cdot (e_l \otimes e_k) \cdot (e_i^{*} \otimes e_j^{*}) \right\}  \nn 
&= \rvec \left \{U_{L^2} \cdot (e_l e_i^{*}) \otimes (e_k e_j^{*}) \right\}  .  
\end{align}
Hence, row $(iv)$ is given by 
\begin{align}\label{c14}
& \rvec \left \{U_{L^2} \sum_{i,j,k,l} E\{\bar{x}_i \bar{x}_l\}  E\{ \bar{x}_j \bar{x}_k\} (e_l e_i^{*}) \otimes (e_k e_j^{*}) \right\} \nn
& \hspace{20ex} = \rvec \left \{ U_{L^2} (R_{X X } \otimes R_{X X }) \right\}. 
\end{align}
Substituting these results into (\ref{c8}) yields (\ref{e9}). 

We next express $K_4(Y)$ in terms of $K_4(X)$ for $Y =AX $. By substituting $X$ with $Y$ in (\ref{e9}) we 
obtain, 
\begin{align}\label{c18}
K_4(Y) &= \mu_4(Y) - \mu_2(Y) \otimes \mu_2(Y) \nn 
&- \rvec \left \{ R_{Y Y } \otimes R_{Y Y } 
+ U_{M^2} \cdot (R_{Y Y } \otimes R_{Y Y }) \right \}.
\end{align}
The first term in the RHS of (\ref{c18}) is given by  (\ref{e13}).
The second term can be expressed using (\ref{e140}) and (\ref{e13}) as,
\begin{align}\label{c19}
\mu_2(Y) \otimes \mu_2(Y) &= [(A \otimes A) \mu_2(X)] \otimes [(A \otimes A) \mu_2(X)] \nn
&= (A \otimes A \otimes  A \otimes A)(\mu_2(X) \otimes \mu_2(X)). 
 \end{align}
For the third term in (\ref{c18}) we use (\ref{e140}) and the well known identity \cite[p. 410]{Lancaster:1985}
\begin{align}\label{c58}
\rvec \left \{A R B^{*} \right\} = (B \otimes A) \rvec\{R\}
\end{align}
to obtain 
\begin{align}\label{c22}
  \rvec & \left \{R_{Y  Y } \otimes R_{Y Y } \right \} x=  \rvec \left \{(A R_{X  X } A^{*}) \otimes (A R_{X  X } A^{*}) \right \} \nn
 &=  \rvec \left \{(A \otimes A) (R_{X  X } \otimes R_{X  X }) (A \otimes A)^{*} \right \} \nn
 &= (A \otimes A \otimes  A \otimes A) \rvec\{ R_{X  X } \otimes R_{X  X }\}.
 \end{align} 
For the last term in (\ref{c18}), we use (\ref{e140}) and the relation 
\begin{align}\label{c23}
U_{M^2} (A \otimes  A) = (A \otimes  A)  U_{L^2}  
\end{align}
to obtain  
\begin{align}\label{c24}
U_{M^2} (R_{Y Y } \! \otimes \! R_{Y Y }) &=
 U_{M^2} (A \! \otimes \! A) (R_{X  X } \! \otimes \!  R_{X  X }) (A \! \otimes \! A)^{*} \nn
&= (A \! \otimes \! A)  U_{L^2} (R_{X  X } \! \otimes \! R_{X  X}) (A \! \otimes \! A)^{*} .
\end{align}
Hence,  from (\ref{c58}), 
\begin{align}\label{c25}
\rvec &  \left \{U_{M^2} (R_{Y Y } \otimes R_{Y Y })\right \} = \nn 
& (A \otimes A \otimes  A \otimes A) \rvec\{ U_{L^2} (R_{X  X } \otimes R_{X  X })\}.
\end{align}
Combining these results we obtain (\ref{e10}). 

Suppose now that the components of $X$ are independent random variables. Then,   
 \begin{align}\label{c27}
 E\{\bar{x}_{l_1} \cdot \bar{x}_{l_2} \cdots \bar{x}_{l_k}\} = 
 \left \{ \begin{array}{cc}
 E\{ \bar{x}_{l}^k\}, & l_1 \! = \! l_2 \! = \! \cdots \! = \! l_k=l \\ 
 0, & \textrm{otherwise}
 \end{array} 
 \right. 
 \end{align}
 and 
 \begin{align}\label{c33}
 \mu_3(X) &= E \left \{ \bar{X}_1 \otimes \bar{X}_1 \otimes \bar{X}_1 \right \} \nn
 &= \sum_{i,j,k} E\{\bar{x}_i \bar{x}_j \bar{x}_k\} (e_i \otimes e_j \otimes e_k ) \nn
&= \sum_{i} E\{\bar{x}_i^3\} (e_i \otimes e_i \otimes e_i ).
 \end{align}
 The central moment $\mu_3(Y)$ follows from (\ref{e13}). Expressing 
 \begin{align}\label{c34}
 A=\sum_{i=1}^L a_i e_i^{*}, 
 \end{align}
we have 
 \begin{align}\label{c35}
 A  \otimes A  \otimes A &= \sum_{i,j,k} (a_i e_i^{*}) \otimes (a_j e_j^{*}) \otimes (a_k e_k^{*}) \nn
 &= \sum_{i,j,k} (a_i \otimes a_j \otimes a_k) (e_i \otimes e_j \otimes e_k )^{*}.
 \end{align}
  Substituting (\ref{c33}) and (\ref{c35}) in (\ref{e13}), and using the orthonormality 
 of the $L^3 \times 1$ vectors $\{e_i \otimes e_j \otimes e_k\}$, we obtain, 
 \begin{align}\label{c36}
 \mu_3(Y) &= \sum_{i=1}^L E\{\bar{x}_i^3\} (a_i \otimes a_i \otimes a_i ) \nn
 &= (A \odot A \odot A) \  \col \left (E \left \{\bar{x}_1^3 \right \},E \left \{\bar{x}_2^3 \right\}, \ldots, E \left \{\bar{x}_L^3 \right\} \right ). 
 \end{align}
 It can similarly 
 be shown that 
 \begin{align}\label{c37}
 \mu_2(Y) &= (A \odot A ) \  \col \left (E \left \{\bar{x}_1^2 \right\},E \left \{\bar{x}_2^2 \right\}, \ldots, E \left \{\bar{x}_L^2 \right\} \right).
 \end{align}
 
When the components of $X$ are independent random variables 
   \begin{align}\label{c31}
 \kappa_4(\bar{x}_i ,\bar{x}_j , \bar{x}_k , \bar{x}_l) = \delta_{ij} \delta_{jk} \delta_{kl} 
 \kappa_4(\bar{x}_i ,\bar{x}_i , \bar{x}_i , \bar{x}_i),  
 \end{align}
and 
  \begin{align}\label{c39}
  \kappa_4(\bar{x}_i, \bar{x}_j, \bar{x}_k, \bar{x}_l) = \left \{
  \begin{array}{cc}
  E\{\bar{x}_i^4\} - 3 (E\{\bar{x}_i^2\})^2, & i \! = \! j \! = \! k \! = \!l \\ 
  0, & \textrm{otherwise.} 
  \end{array} 
  \right . 
   \end{align} 
Substituting (\ref{c39}) into (\ref{c8}) yields,  
   \begin{align}\label{c40}
   K_4(X) &= \sum_{i,j,k,l} \kappa_4(\bar{x}_i, \bar{x}_i, \bar{x}_i, \bar{x}_i) \delta_{ij} \delta_{jk} \delta_{kl} (e_i \otimes e_j \otimes e_k \otimes e_l) \nn
   &= \sum_p \kappa_4(\bar{x}_p, \bar{x}_p, \bar{x}_p, \bar{x}_p) (e_p \otimes e_p \otimes e_p \otimes e_p) .
   \end{align}
To find $K_4(Y)$ we use (\ref{e10}) where 
\begin{align}\label{c41}
A \otimes A \otimes  A \otimes A = \!
   \sum_{i,j,k,l} \! (a_i  \! \otimes \!  a_j \!  \otimes\!  a_k \otimes \! a_l) 
     (e_i \! \otimes \! e_j \! \otimes \! e_k \! \otimes e_l)^{*}
    \end{align}
   as follows from (\ref{c34})-(\ref{c35}), and $K_4(X)$ is given in (\ref{c40}). Using 
   orthonormality of the $L^4 \times 1$ vectors $\{e_i \otimes e_j \otimes e_k \otimes e_l\}$
   as in (\ref{c36}), yields, 
   \begin{align}\label{c42}
   K_4(Y) & = \sum_{p=1}^L \kappa_4(\bar{x}_p, \bar{x}_p, \bar{x}_p, \bar{x}_p) (a_p \otimes a_p \otimes a_p \otimes a_p) \nn
 &= (A \odot A \odot A \odot A) \left ( \begin{array}{c}
 \kappa_4(\bar{x}_1, \bar{x}_1, \bar{x}_1, \bar{x}_1) \\ 
 \kappa_4(\bar{x}_2, \bar{x}_2, \bar{x}_2, \bar{x}_2) \\ 
 \vdots \\ 
 \kappa_4(\bar{x}_L, \bar{x}_L, \bar{x}_L, \bar{x}_L) 
 \end{array} \right ),   
   \end{align}
   which is (\ref{e16}). 
We conjecture that the same relation holds for all higher order cumulants. 

When the components of $X$ are independent Poisson random variables
with $E\{X\}=\lambda$,  
\begin{align}\label{c43}
 E\{\bar{x}_i^2\} &= E\{\bar{x}_i^3\} = \lambda_i \nn
 E\{\bar{x}_i^4\} & = \lambda_i + 3 \lambda_i^2 \nn
 \kappa_4(\bar{x}_i, \bar{x}_i, \bar{x}_i, \bar{x}_i) &= (\lambda_i + 3 \lambda_i^2) - 3 \lambda_i^2 = \lambda_i,  
 \end{align} 
 and (\ref{e7}) follows.

\bibliographystyle{IEEEtran}
\bibliography{IEEEabrv,CumulantsRef}
\end{document}